\documentclass[12pt]{article}

\usepackage{epsf}
\usepackage[dvips]{graphicx}

\begin{document}

\begin{center}
{\bfseries SINGLE AND DOUBLE PION PRODUCTION IN NP COLLISIONS AT 1.25 GEV WITH HADES.}
%{\bfseries Single and double pion production in np collisions at 1.25 GeV with HADES.}

\vskip 5mm
A.K.~Kurilkin$^{1\star}$, 
G.~Agakishiev$^{\star}$, 
A.~Balanda$^{\&}$, 
D.~Belver$^{\uplus}$,
A.~Belyaev$^{\star}$,
A.~Blanco$^{\$}$, 
M.~B\"{o}hmer$^{\bullet}$, 
J.~L.~Boyard$^{\cap}$, 
P.~Cabanelas$^{\uplus}$, 
E.~Castro$^{\uplus}$, 
S.~Chernenko$^{\star}$, 
J.~D\'{\i}az$^{\uplus}$,  
A.~Dybczak$^{\&}$, 
E.~Epple$^{\bullet}$, 
L.~Fabbietti$^{\bullet}$, 
O.~Fateev$^{\star}$, 
P.~Finocchiaro$^{\%}$, 
P.~Fonte$^{\$,a}$, 
J.~Friese$^{\bullet}$, 
I.~Fr\"{o}hlich$^{\mp}$, 
T.~Galatyuk$^{\mp}$, 
J.~A.~Garz\'{o}n$^{\uplus}$, 
A.~Gil$^{\sqcap}$, 
M.~Golubeva$^{\circ}$, 
D.~Gonz\'{a}lez-D\'{\i}az$^{+}$, 
F.~Guber$^{\circ}$, 
T.~Hennino${\cap}$, 
R.~Holzmann$^{+}$, 
P.~Huck$^{\bullet}$,
A.~Ierusalimov$^{\star}$, 
I.~Iori$^{\otimes,c}$,
A.~Ivashkin$^{\circ}$,
M.~Jurkovic$^{\bullet}$,
B.~K\"{a}mpfer$^{\pm,b}$,
T.~Karavicheva$^{\circ}$,
I.~Koenig$^{+}$,
W.~Koenig$^{+}$,
B.~W.~Kolb$^{+}$,
A.~Kopp$^{\times}$,
G.~Korcyl$^{\&}$,
GK~Kornakov$^{\uplus}$,
R.~Kotte$^{\pm}$,
A.~Kozuch$^{\&,d}$,
A.~Kr\'{a}sa$^{\cup}$,
F.~Krizek$^{\cup}$,
R.~Kr\"{u}cken$^{\bullet}$,
H.~Kuc$^{\&}$,
W.~K\"{u}hn$^{\times}$,
A.~Kugler$^{\cup}$,
A.~Kurepin$^{\circ}$,
P.~Kurilkin$^{\star}$,
P.~KŠhlitz$^{\pm}$,
V.~Ladygin$^{\star}$,
J.~Lamas-Valverde$^{\uplus}$,
S.~Lang$^{+}$,
K.~Lapidus$^{\circ}$,
T.~Liu$^{\bullet}$,
L.~Lopes$^{\$}$,
M.~Lorenz$^{\mp}$,
L.~Maier$^{\bullet}$,
A.~Mangiarotti$^{\$}$,
J.~Markert$^{\mp}$,
V.~Metag$^{\times}$,
B.~Michalska$^{\&}$,
J.~Michel$^{\mp}$,
C.~M\"{u}ntz$^{\mp}$,
L.~Naumann$^{\pm}$,
Y.~C.~Pachmayer$^{\mp}$,
M.~Palka$^{\mp}$,
Y.~Parpottas$^{\div}$,
V.~Pechenov$^{+}$,
O.~Pechenova$^{\mp}$,
J.~Pietraszko$^{\mp}$,
W.~Przygoda$^{\&}$,
B.~Ramstein$^{\cap}$,
A.~Reshetin$^{\circ}$,
J.~Roskoss$^{\times}$,
A.~Rustamov$^{+}$,
A.~Sadovsky$^{\circ}$,
P.~Salabura$^{\&}$,
A.~Schmah$^{\bullet}$,
J.~Siebenson$^{\bullet}$,
Yu.~G.~Sobolev$^{\cup}$,
S.~Spataro$^{\times,e}$,
H.~Str\"{o}bele$^{\mp}$,
J.~Stroth$^{\mp,+}$,
C.~Sturm$^{+}$,
M.~Sudol$^{\cap}$,
A.~Tarantola$^{\mp}$,
K.~Teilab$^{\mp}$,
P.~Tlusty$^{\cup}$,
M.~Traxler$^{+}$,
R.~Trebacz$^{\&}$,
H.~Tsertos$^{\div}$,
T.~Vasiliev$^{\star}$,
V.~Wagner$^{\cup}$,
M.~Weber$^{\bullet}$,
J.~W\"{u}stenfeld$^{\pm}$,
S.~Yurevich$^{+}$,
Y.~Zanevsky$^{\star}$

\vskip 5mm

{\small
$^{\%}$ {\it Istituto Nazionale di Fisica Nucleare - Laboratori Nazionali del Sud, 95125~Catania, Italy } \\
$^{\$}$ {\it LIP-Laborat\'{o}rio de Instrumenta\c{c}\~{a}o e F\'{\i}sica Experimental de Part\'{\i}culas , 3004-516~Coimbra, Portugal } \\
$^{\&}$ {\it Smoluchowski Institute of Physics, Jagiellonian University of Cracow, 30-059~Krak\'{o}w, Poland} \\
$^{+}$ {\it GSI Helmholtzzentrum f\"{u}r Schwerionenforschung GmbH, 64291~Darmstadt, Germany } \\
$^{\pm}$  {\it Institut f\"{u}r Strahlenphysik, Forschungszentrum Dresden-Rossendorf, 01314~Dresden, Germany } \\
$^{\star}$  {\it Joint Institute of Nuclear Research, 141980~Dubna, Russia } \\
$^{\mp}$  {\it Institut f\"{u}r Kernphysik, Goethe-Universit\"{a}t, 60438 ~Frankfurt, Germany } \\
$^{\times}$  {\it II.Physikalisches Institut, Justus Liebig Universit\"{a}t Giessen, 35392~Giessen, Germany } \\
$^{\otimes}$  {\it Istituto Nazionale di Fisica Nucleare, Sezione di Milano, 20133~Milano, Italy } \\
$^{\circ}$  {\it Institute for Nuclear Research, Russian Academy of Science, 117312~Moscow, Russia } \\
$^{\bullet}$  {\it Physik Department E12, Technische Universit\"{a}t M\"{u}nchen, 85748~M\"{u}nchen, Germany } \\
$^{\div}$  {\it Department of Physics, University of Cyprus, 1678~Nicosia, Cyprus } \\
$^{\cap}$ {\it Institut de Physique Nucl\'{e}aire (UMR 8608), CNRS/IN2P3 - Universit\'{e} Paris Sud, F-91406~Orsay Cedex, France } \\
$^{\cup}$  {\it Nuclear Physics Institute, Academy of Sciences of Czech Republic, 25068~Rez, Czech Republic } \\
$^{\uplus}$  {\it Departamento de F\'{\i}sica de Part\'{\i}culas, Univ. de Santiago de Compostela, 15706~Santiago de Compostela, Spain } \\
$^{\sqcap}$  {\it Instituto de F\'{\i}sica Corpuscular, Universidad de Valencia-CSIC, 46971~Valencia, Spain } \\
$^{a}$  {\it Also at ISEC Coimbra, ~Coimbra, Portugal } \\
$^{b}$  {\it Also at Technische Universit\"{a}t Dresden, 01062~Dresden, Germany } \\
$^{c}$  {\it Also at Dipartimento di Fisica, Universit\`{a} di Milano, 20133~Milano, Italy } \\
$^{d}$  {\it Also at Panstwowa Wyzsza Szkola Zawodowa , 33-300~Nowy Sacz, Poland } \\
$^{e}$  {\it Also at Dipartimento di Fisica Generale, Universit\'{a} di Torino, 10125 ~Torino, Italy } \\
$\dag$ {\it
E-mail: akurilkin@jinr.ru
}}
\end{center}

\vskip 5mm

\begin{center}
\begin{minipage}{150mm}
\centerline{\bf Abstract}
The preliminary results on charged pion production in np collisions at an incident beam energy of 1.25 GeV measured with HADES are presented. The np reactions were isolated in dp collisions at 1.25 GeV/u using the Forward Wall hodoscope, which allowed to register spectator protons. The results for $np \rightarrow pp\pi^{-}$, $np \rightarrow np\pi^{+}\pi^{-}$ and $np \rightarrow d\pi^{+}\pi^{-}$ channels are compared with OPE calculations. 
A reasonable agreement between experimental results and the predictions of the OPE$+$OBE model is observed.
\end{minipage}
\end{center}

\vskip 10mm

\section{Introduction}

\hspace{0.5cm}
Pion production in nucleon-nucleon collisions has been the subject of considerable interest in nuclear and particle physics for many years. A number of experiments have been performed, spanning the energy region from threshold to many GeV's\cite{Shimizu_pp,Tsuboyama}. The bulk of the experimental data has come from pp collision. The pn interaction data in the low and medium energy regions are scare despite of their importance not only for understanding the NN interactions but also for the interpretation of medium-energy heavy-ion interactions.
The np interactions are studied by using deuteron-proton (dp) collisions with the deuteron either as the projectile or as the target. The main reason for this situation is due to the difficulty to create pure monoenergetic neutron beams.

%These interactions, in particular the inelastic interactions, are good examples of a few body problem.   

One of the questions which is not answered at the present time is the contribution of isoscalar (I=0) partial waves to the inelastic np collision.
The neutron-proton scattering amplitude contains both isoscalar (I=0) and isovector (I=1) parts and, while the isovector part is rather well known, even the order of magnitude of the total isoscalar cross section is badly determined. Usually, this cross-section is extracted from the difference of the total cross-sections of the pion production reactions: $np \rightarrow pp\pi^{-}$ and $pp \rightarrow pp\pi^{0}$.

%Study of np reactions is necessary in order to obtain the contribution of isoscalar (I=0) partial waves to the inelastic np collision.
%%The first one is that the np reaction is important to obtain the contribution of isoscalar (I=0) partial waves to the inelastic np collision. 
%The neutron-proton scattering amplitude contains both isoscalar (I=0) and isovector (I=1) parts, and while the isovector part is rather well known, even the order of magnitude of the total isoscalar cross sections is badly determined. Usually, this cross-section is extracted from the difference of the total cross-sections of the pion production reactions: $np \rightarrow pp\pi^{-}$ and $pp \rightarrow pp\pi^{0}$.    

A study of the double pion production in the NN collisions is one way to obtain information about nucleon-nucleon, pion-nucleon, pion-pion interactions and for the investigation of resonances properties. The comparison of the double-pion production from np and pp interaction can bring new constraints on the recently reported $e^{+}e^{-}$ excess in the np reaction \cite{hades_enhancement}. 
Also the study of single and double pion production in np collisions at different energies is important for the determination of both the energy dependence of the total np cross section and the contribution of inelastic channels to np interactions.
This paper presents preliminary results for the $np \rightarrow pp\pi^{-}$, $np \rightarrow np\pi^{+}\pi^{-}$ and $np \rightarrow d\pi^{+}\pi^{-}$ channels at 1.25 GeV.

%Pion production in nucleon-nucleon collisions has been the subject of considerable interest in nuclear and particle physics for many years. A number of experiments have been performed, spanning the energy region from theshold to many GeV. The bulk of the experimental data has come from pp collision, due to the difficulty in obtaining intense and monoenergetic neutron beams. Despite the importance of the p-n interactions data in the low and medium energy regions not only for understanding of the NN interactions but also for the interpretation of medium energy heavy ion interactions, there are still not enough data on neutron-proton(np) interactions.

%Data on low energy p-n interactions are important as well for understanding of the NN interactions as for interpretation of medium energy heavy ion collisions. 
%However, the bulk of the experimental data has come from pp collision, due to the difficulty in obtaining intense and monoenergetic neutron beams.  
%Despite the importance of the nucleon-nucleon (NN) interaction data in the low and medium energy regions not only for understanding of the NN interactions but also for the interpretation of medium energy heavy ion interactions, there are still not enough data on neutron-proton(np) interactions. 

\section{High Acceptance Di-Electron Spectrometer}
\hspace{0.5cm} The High Acceptance Di-Electron Spectrometer(HADES) is an unique apparatus installed at the heavy-ion synchrotron SIS18 at GSI Darmstadt \cite{hades1}. 
It is designed for high-resolution and high-acceptance dielectron spectroscopy in hadron-hadron, hadron-nucleus and nucleus-nucleus reactions at beam energies in the range from 1A GeV to 2A GeV. 
The major part of the HADES physics program focuses on in-medium properties of the light vector mesons, $\rho$, $\omega$ and $\phi$. 

The HADES spectrometer consists of 6 identical sectors covering the full azimuthal angle and polar angles from $18^\circ$ to $85^\circ$ relative to beam direction. Each sector of the spectrometer contains a Ring Imaging Cherenkov Detector (RICH) operating in a magnetic field, inner Multi Wire Drift Chambers (MDCs) in front of magnetic field, outer MDCs behind the magnetic field, Time Of Flight (TOF and TOFino) detectors and a electromagnetic cascade detector (Pre-Shower). A detailed description of the HADES spectrometer can be found in \cite{hades1}.   
 Momentum measurement for charged particles is achieved by tracking the particles in front of and behind a toroidal field generated by six superconducting coils arranged around the beam axis. A hadron-blind Ring Imaging CHerenkov detector (RICH) placed around the target region is used for electron identification, together with TOF/TOFINO and an electromagnetic pre-shower detector (Pre-Shower). Particle identification is also provided using the correlations between time-of-flight and momentum of charged pions, protons and deuterons. 
Forward Wall (FW) scintillator hodoscope covering the polar angle between $1^{\circ}$ and $7^{\circ}$ was installed lately 2007 for tagging the spectator proton in $dp$ reactions.    
With technical design features, HADES can obtain data with high quality and statistical significance.

\begin{figure}[ht]
 \centerline{
 \includegraphics[width=8cm,height=8cm]{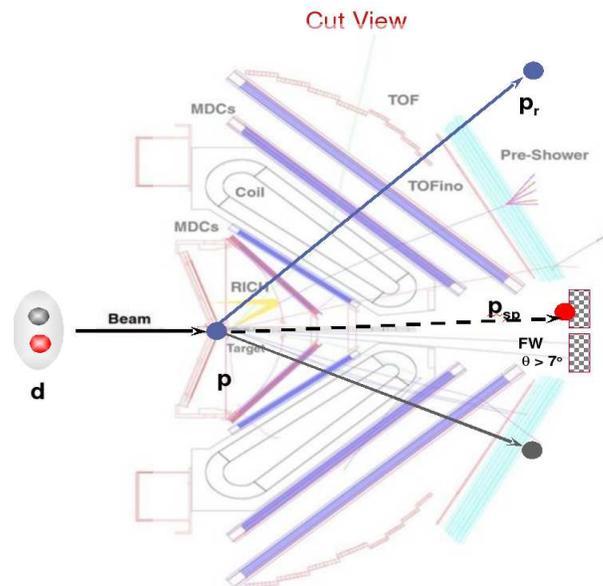}}
 \caption{Cut through two sectors of the HADES spectrometer. The magnet coils are projected onto the cut plane to visualize the toroidal magnetic field. A schematic view of the quasi-free $n+p$ reaction is shown. }
\end{figure}

\section{dp experiment}

%The experimental run was performed with the High Acceptance Di-Electron Spectrometer\cite{hades1}
%[5] 
%installed at the GSI, Germany. 
A deuteron beam of $10^7$ particles/s with kinetic energy of 1.25 AGeV was incident on a liquid hydrogen cell with a length of 5 cm, corresponding to a interaction length of $\rho$d=0.35 $g/cm^2$.   
Quasi-free $n-p$ reactions were selected at the trigger level by detection of fast spectator protons from the deuterium break-up in the FW. The detection of the spectator protons by the FW allowed to suppress the contribution of quasi-free $p-p$ reactions \cite{Lapidus_FW}. 
%[17].
The FW is an array which consists of nearly 300 scintillating cells with each 2.54 cm thickness. During the dp experiment it was located 7 m downstream the target.
The estimated time resolution of the FW is about 500 ps; thus the estimated momentum resolution of the detected particles (protons) is $\sim 11\%$.

\section{Results}

Fig.~2 exhibits the preliminary missing mass and invariant mass spectra for the np$\rightarrow$$pp\pi^{-}$ channel. Both spectra are not efficiency corrected.
The vertical line in Fig. 2.a) at 0.03 $GeV^{2}/c^{4}$ shows the criterion on the $M^{2}_{miss}$ which was applied to remove events with an additional $\pi^{0}$ production. 
The experimental $M_{inv}$ distribution is shown in Fig.1.b). Two possible proton combinations were taken into account.
The maximum of the distribution correspond to the $\Delta(1232)$ resonance.    

%Events with the additional $\pi^{0}$ production were removed by applying the cut on squared missing mass(vertical line on Fig.2.a).
%$M_{inv}$ spectra of $p\pi^{-}$ pair for the simulation in the PLUTO are presented in Fig.2.b). Open and filled histograms correspond to all events and to such ones within the HADES acceptance.

\vspace{-0.5cm}

\begin{figure}[h]
 \centerline{
 \includegraphics[width=12cm,height=6.2cm]{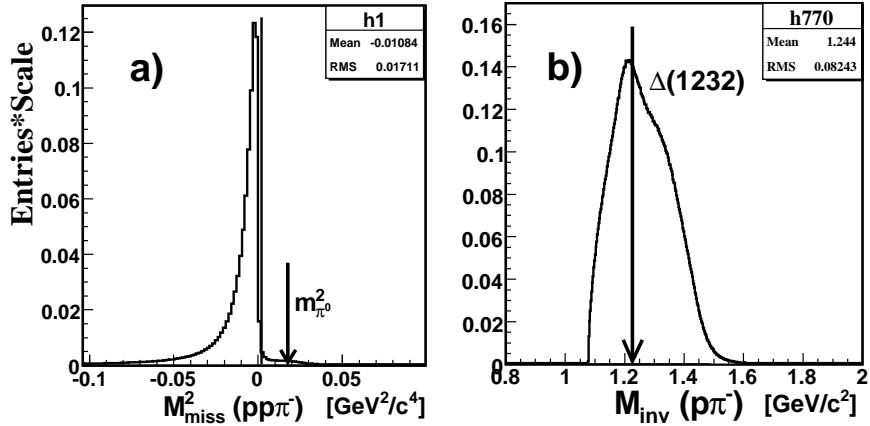}}
 \caption{Preliminary results for np$\rightarrow$$pp\pi^{-}$ channel. a) - the $pp\pi^{-}$ missing mass spectrum and b) the $p\pi^{-}$ invariant mass spectrum. The dashed arrow in a) and the solid arrow in b) correspond to the squared mass of $\pi^{0}$ and the mass of $\Delta(1232)$ resonance, respectively.
\label{f1} }
\end{figure}

Figs. 3. and 4. exhibit preliminary spectra of $M^{2}_{miss}$ and $M_{inv}$ for the np$\rightarrow$$d\pi^{+}\pi^{-}$ and np$\rightarrow$$np\pi^{+}\pi^{-}$ channels, respectively. 
The spectra are not efficiency corrected. Distributions of $M^{2}_{miss}$ in Fig. 3.a) and Fig. 4.a) show that channels are separated correctly.
Fig. 3.b) and 3.c) correspond to the $M_{inv}$ spectra of $\pi^{+}\pi^{-}$ and $d\pi^{}$. Fig. 4.b), 4.c), 4.d), 4.e), 4.f) correspond to the $M_{inv}$ spectra of $\pi^{+}\pi^{-}$, $p\pi^{+}$, $p\pi^{-}$, $n\pi^{+}$, $n\pi^{-}$, respectively. $M_{inv}$ spectra of $p\pi^{+}$ in Fig. 4.c) show that the np reaction should follow via np$\rightarrow$$\Delta^{++}n\pi^{-}\rightarrow$$np\pi^{+}\pi^{-}$.

\begin{figure}[h]
 \centerline{
 \includegraphics[width=15cm,height=7.cm]{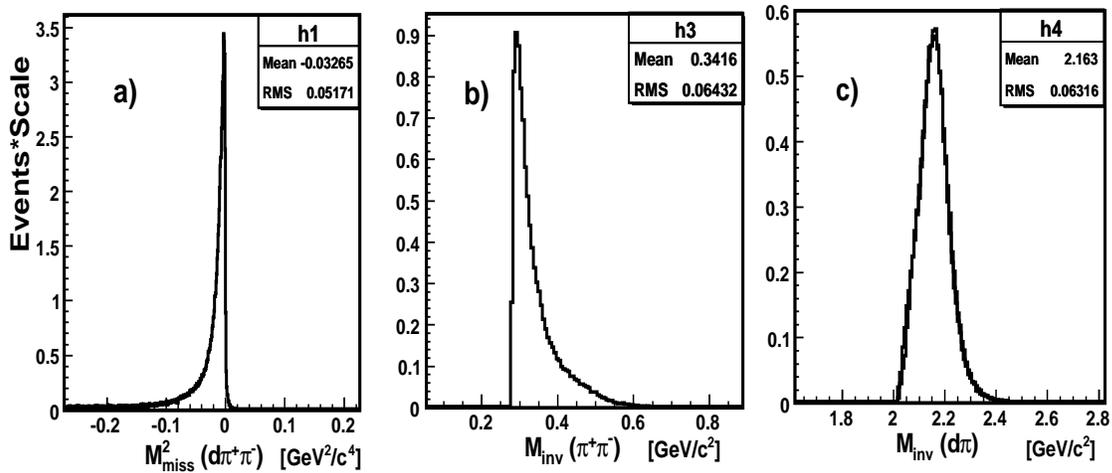}}
 \caption{Preliminary results for np$\rightarrow$$d\pi^{+}\pi^{-}$ channel. a) - the $d\pi^{+}\pi^{-}$ missing mass spectra, b) - the $\pi^{+}\pi^{-}$ invariant mass spectra and c) - $d\pi^{}$ invariant mass spectra.  \label{f2}}
\end{figure}

\begin{figure}[h]
 \centerline{
 \includegraphics[width=15cm,height=8.cm]{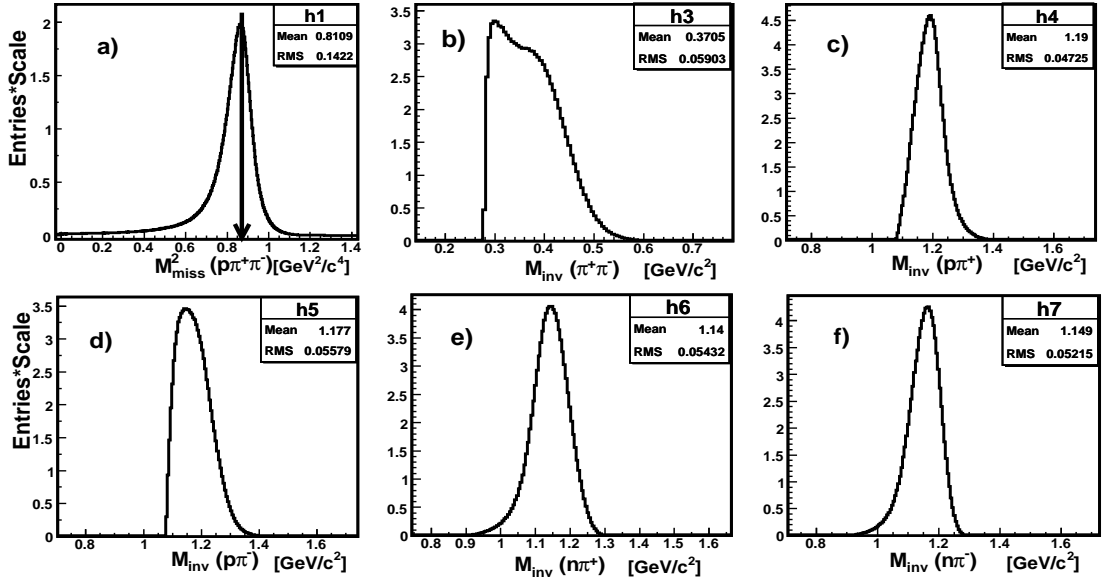}}
 \caption{Preliminary results for np$\rightarrow$$np\pi^{+}\pi^{-}$ channel. a) the $p\pi^{+}\pi^{-}$ - missing mass spectra, b) the $\pi^{+}\pi^{-}$ - invariant mass spectra, c) the $p\pi^{+}$ - invariant mass spectra, d) the $p\pi^{-}$ - invariant mass spectra, e) the $n\pi^{+}$ - invariant mass spectra, f) the $n\pi^{-}$ - invariant mass spectra. The solid arrow in a) corresponds to the squared neutron mass.  \label{f3}}
\end{figure}

\section{Prediction of the Pluto and OPE$+$OBE models}

The predictions of Pluto event generator \cite{Lapidus} for the np$\rightarrow$$pp\pi^{-}$ and np$\rightarrow$$d\pi^{+}\pi^{-}$ channels are presented in Fig. 5.
The spectra show events which belong to the geometric acceptance of HADES.
%The predictions of Pluto model for the  np$\rightarrow$$pp\pi^{-}$ and np$\rightarrow$$d\pi^{+}\pi^{-}$ channels are presented in Fig.5. 
%Fig. 5.a) corresponds to the $M_{inv}$($p\pi^{-}$) spectrum, taking into account two possible proton combinations. 
The excitation of $\Delta(1232)$ resonance only was included for the np$\rightarrow$$pp\pi^{-}$ process.  
%Fig. 5.b) and 5.c) correspond to the $M_{inv}$($\pi^{+}\pi^{-}$) and $M_{inv}$($d\pi^{}$) spectra for the np$\rightarrow$$d\pi^{+}\pi^{-}$ channel, respectively.
The phase space was generated for the np$\rightarrow$$d\pi^{+}\pi^{-}$ channel. 
The behavior of $M_{inv}$ spectra is qualitatively reproduced by Pluto event generator for both channels. 
However the $\Delta(1232)$ resonance is not so clearly visible in the not corrected by the efficiency experimental distribution (see Fig. 2.b) ).   
%However the experimental data require the correction on the efficiency.  

\begin{figure}[ht]
 \centerline{
 \includegraphics[width=16cm,height=6.5cm]{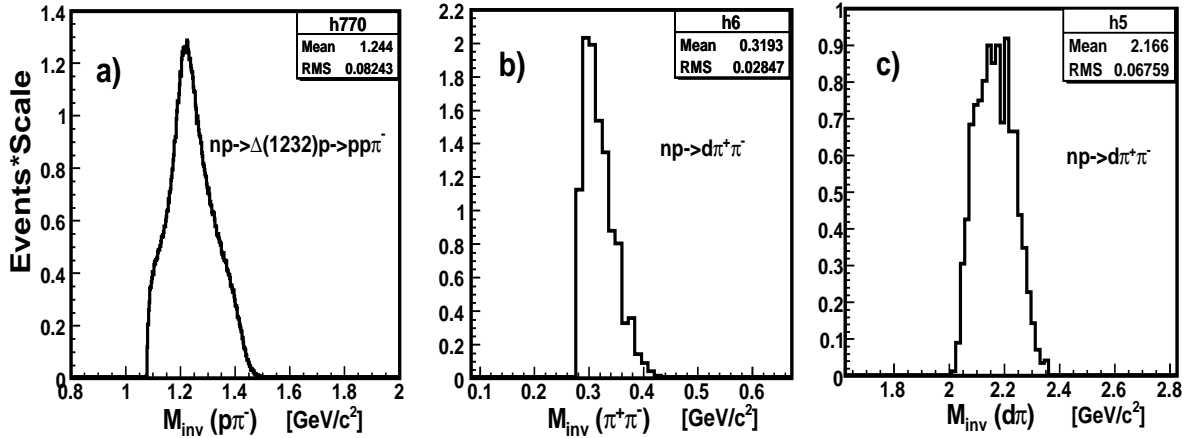}}
 \caption{Prediction of Pluto for the np$\rightarrow$$pp\pi^{-}$ and np$\rightarrow$$d\pi^{+}\pi^{-}$ channels. Explanations are given in the text.\label{f4}}
\end{figure}

\begin{figure}[ht]
 \centerline{
 \includegraphics[width=14cm,height=8.cm]{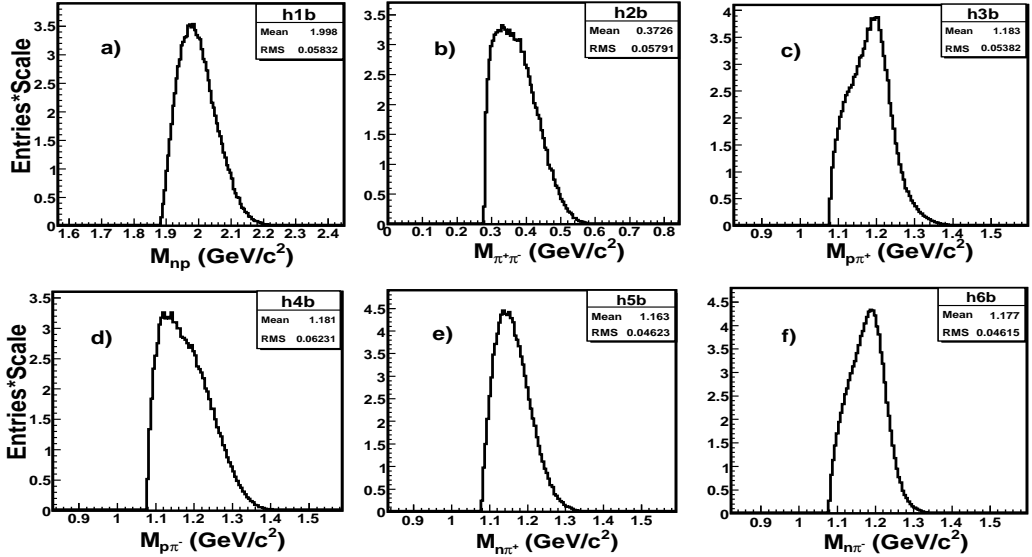}}
 \caption{Predictions of OPE$+$OBE model calculations for np$\rightarrow$$np\pi^{+}\pi^{-}$ channel. \label{f5}}
\end{figure}

Fig. 6 exhibits the prediction of OPE$+$OBE (one $\pi$-exchange and one barion-exchange) model \cite{ierusalimov} calculations for the $np \rightarrow np\pi^{+}\pi^{-}$ channel. This model include also the interference of the main diagrams of reggeized one $\pi$-exchange model (OPER) for $np \rightarrow np\pi^{+}\pi^{-}$ reaction at the energies 1-5 GeV \cite{ierusalimov}. 
%Fig. 6 a), b), c), d), e), f) correspond to the $M_{inv}$($np$), $M_{inv}$($\pi^{+}\pi^{-}$), $M_{inv}$($p\pi^{+}$), $M_{inv}$($p\pi^{-}$), $M_{inv}$($n\pi^{+}$), $M_{inv}$($n\pi^{-}$) spectra, respectively. 
%Invariant mass spectra of $p\pi^{+}$ and $n\pi^{-}$  clearly show an excitation of $\Delta^{++}$ and $\Delta^{-}$ resonances.  
The behavior of $M_{inv}$ spectra (see Fig. 4) is qualitatively reproduced by predictions of the OPE$+$OBE model. However the experimental data require the correction on the efficiency.  

%Although the experimental data are not corrected by the efficiency, the behavior of $M_{inv}$ spectra is qualitatively reproduced by predictions of the OPE$+$OBE model.
%The reasonable agreement between preliminary experimental results and the predictions of the OPE$+$OBE model is observed. 
\vspace{-0.5cm}

\section{Conclusion}
\hspace{0.5cm}  The preliminary results on charged pion production in np collisions at 1.25 GeV obtained with HADES are presented. 
%The preliminary results on charged pion production studied in np collisions at 1.25 GeV with the HADES spectrometer are presented. These results will serve as entry for further analysis steps. 
A reasonable agreement of preliminary experimental results and predictions of the Pluto event generator and OPE$+$OBE model is observed.
%for the $np \rightarrow np\pi^{+}\pi^{-}$ channel 
More detailed information will be obtained through a comparison the efficiency corrected experimental spectra and the simulation results.

The work has been supported in part by the Russian Foundation for Basis Research grant $ No.$ 10-02-00087a.

\end{document}